\documentclass[prl, twocolumn, showpacs, showkeys, preprintnumbers, amsmath, amssymb]{revtex4-1}

\usepackage{graphicx}

\begin{document}

\author{A.~S.~Zhuravlev}
\affiliation{Institute of Solid State Physics, Russian Academy of Sciences, Chernogolovka, 142432 Russia}
\author{V.~A.~Kuznetsov}
\altaffiliation[Also at ]{Moscow Institute of Physics and Technology, Dolgoprudniy, 141700 Russia}
\email{volod_kuzn@iccp.ac.ru}
\affiliation{Institute of Solid State Physics, Russian Academy of Sciences, Chernogolovka, 142432 Russia}
\author{V.~E.~Bisti}
\affiliation{Institute of Solid State Physics, Russian Academy of Sciences, Chernogolovka, 142432 Russia}
\author{L.~V.~Kulik}
\affiliation{Institute of Solid State Physics, Russian Academy of Sciences, Chernogolovka, 142432 Russia}
\author{V.~E.~Kirpichev}
\affiliation{Institute of Solid State Physics, Russian Academy of Sciences, Chernogolovka, 142432 Russia}
\author{I.~V.~Kukushkin}
\affiliation{Institute of Solid State Physics, Russian Academy of Sciences, Chernogolovka, 142432 Russia}
\author{S.~Schmult}
\affiliation{TU Dresden, Institute of Semiconductors and Microsystems, N{\"o}thnitzer Stra{\ss}e 64, 01187 Dresden, Germany}

\title{Artificially constructed plasmaron and plasmon-exciton molecule in 2D-metals}

\keywords{plasmaron}

\date{\today}

\begin{abstract}
Resonant optical excitation was used to create a macroscopic non-equilibrium ensemble of 'dark' excitons with unprecedentedly long lifetime in a two-dimensional (2D) electron system  placed in a quantizing magnetic field. Exotic three-particle and four-particle states, plasmarons and plasmon-exciton molecules, coupled with the surrounding electrons through the collective plasma oscillations are engineered. Plasmarons and plasmon-exciton molecules are manifested as new features in the recombination spectra of non-equilibrium systems.
\end{abstract}

\maketitle

Convincing evidence has been acquired regarding the existence of three-particle excitations participating in collective plasma oscillations in metals, \textit{i.e.}, plasmarons. Plasmarons were predicted as far back as in 1967~\cite{hedin1967}. However, the first unambiguous demonstration of their existence has been presented only recently. Plasmaron peaks were observed near the Dirac point in the angle-resolved photoemission spectra of graphene~\cite{bostwick2010}. Plasmaron features were also seen in GaAs/AlGaAs quantum wells (QWs) by means of pulsed tunneling spectroscopy~\cite{dial2012}.
There have also been few reports on plasmarons in 3D-metals~\cite{tediosi2007}. The plasmaron is essentially a specific example of a more general class of charged excitations in metals, trions, which have been extensively discussed with regard to recombination spectra of two-dimensional electron systems~\cite{shay1971, shields1995}. The main issue of trion physics is related to the possibility of observing the recombination of free (translation invariant) trions~\cite{dzyubenko2000}. In the work presented here, we show that a trion, a plasmaron, as well as a more intricate four-particle plasmon-exciton molecule, form naturally in 2D-metal placed in a quantizing magnetic field.

Before discussing the possibility of observing 2D-plasmarons, it is necessary to state the problem of creating 2D translation-invariant two-particle exciton states in metals.
These states are likely to form near the Dirac point in graphene owing to a significant reduction of the electron-hole masses~\cite{bostwick2010}.
In the case of massive systems, the formation of exciton states becomes more intricate owing to screening of the excitons by free carriers~\cite{cui} or localisation of excitons in the random potential~\cite{shields1995}.
Even if this is achieved, a question arises whether the third particle attached to the two-particle state would manifest itself in the optical spectra.
A photon is unable to create a charged three-particle excitation, and a translationary invariant  three-particle excitation cannot decay with photon emission.
The influence of the third particle can be visualized through inner transitions of a three-particle state only~\cite{dzyubenko2000}. Thus, to observe plasmaron features in optical spectra, one has to create a system of free excitons with lifetime long enough to catch an extra particle, then an inner transition should occur with emitting a photon, and finally, the emitted photon should fall in a frequency range accessible for its reliable detection. It turns out that all those conditions are well achieved for a high mobility electron gas confined in a GaAs/AlGaAs quantum well.

As was already mentioned, in metals such as 2D-electron gas, the excitons are unstable~\cite{cui}.
However, in quantizing magnetic fields and at temperatures well below the cyclotron quantization energy, a 2D-metal turns into a special type of 2D-insulator at even integer quantum Hall states~\cite{klitzing1980}. Excitations in quantum Hall insulators are magnetoexcitons. They combine properties of excitons in 2D-insulators and plasmons in 2D-metals~\cite{bychkov1981}.
The simplest realizations are magnetoexcitons formed by an electron promoted from the occupied zeroth Landau level to the empty first Landau level and by the vacancy left in the Fermi sea of electrons. There are two magnetoexcitons: a spin-singlet with total spin $S = 0$; and a spin-triplet with total spin $S = 1$ and spin projections along the magnetic field axis $S_z=-1, 0, 1$~\cite{kulik2005}. The singlet magnetoexciton $S = 0$ is essentially the Kohn magnetoplasmon~\cite{kohn1961}.
Its fast relaxation to the ground state occurs via dipole cyclotron radiation~\cite{zhang2014}.
In contrast to the spin-singlet, the spin-triplet magnetoexciton is not radiatively active owing to electron spin conservation.
It is a 'dark' magnetoexciton. It interacts with all other electrons under the Fermi level. The many-body Coulomb interaction lowers its energy below the cyclotron energy.
Thus, the spin-triplet magnetoexciton is the lowest energy excited state in the system~\cite{kulik2005}. Because of this and the spin conservation, the spin-triplet magnetoexcitons exhibit extremely slow relaxation with the relaxation time reaching hundreds of microseconds~\cite{kulik2015}.

An experimental technique for the creation, manipulation, and detection of macroscopic ensembles of spin-triplet magnetoexcitons has been recently developed~\cite{kulik2015}.
The ensemble can be created with a resonant optical excitation (the resonances appear due to the orbital and the spin quantization), whereas the usage of extra non-resonant photo-pumping adds high-energy electron-hole pairs to the ensemble. During relaxation, the latter may dissociate and relax to the ground state.
Yet, depending on the amount of extra high-energy pairs and the density of spin-triplet magnetoexcitons, some of the dissociated valence holes stick to the magnetoexcitons as extra particles with the formation of positively charged three-particle states, whereas some non-dissociated pairs stick to the magnetoexcitons with the formation of neutral four-particle states.
Negatively charged three-particle states are also likely to form, but since the extra electron settles in the conductance band, they are not optically active in the visible spectral range and therefore are not discussed here. Thus, by manipulating the densities of resonant and non-resonant excitations, one can artificially construct all possible stable states consisting of spin-triplet magnetoexcitons, electrons, and holes.

To create a macroscopic ensemble of spin-triplet excitons, we utilized two sets of high-quality heterostructures (dark mobility is in the range of $5\!-\!20\times 10^6\,$cm${}^2$/Vs) with symmetrically doped GaAs/AlGaAs single quantum wells of two widths: 17 and 35 nm. The samples were placed into a $^3$He pumped cryostat inserted in a superconducting solenoid.
A continuous wave tuneable laser with a narrow spectral line was employed as an optical source for resonant excitation of spin-triplet magnetoexcitons. A more broad band optical source was used to add extra electron-hole pairs in the system. In fact, similar to resonant excitations, non resonant excitations also provide both spin-triplet magnetoexcitons and free electron-hole pairs. The difference is in their relative amounts. The pumping efficiency for spin-triplet magnetoexcitons varies in orders of magnitude with small (around 0.1 nm) detuning of the laser wavelength from resonance.

\begin{figure}
\includegraphics{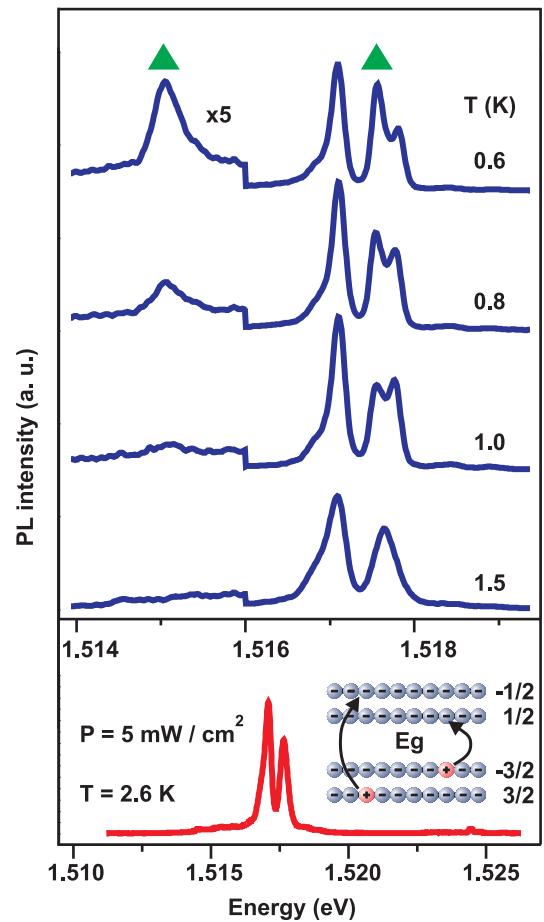}
\caption{\label{fig:fig1}
Bottom: PL spectrum at 2.6 K, at which the 'dark' spin-triplet magnetoexcitons are ionized.
The inset shows the allowed optical transitions.
Top: PL spectra at various temperatures measured at a pumping power density of 5 mW/cm$^2$.}
\end{figure}

\begin{figure}
\includegraphics{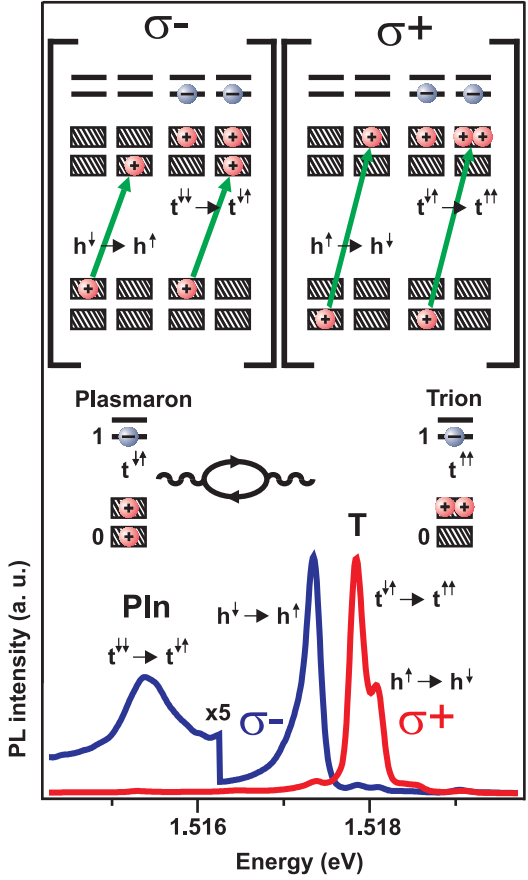}
\caption{\label{fig:fig2}
Top: scheme showing allowed optical transitions in $\sigma_{-}$ and $\sigma_{+}$ polarizations for single and three particle states.
The letters h and t denote the single-particle and trion states, respectively.
The arrows denote the z-projections of the spin for holes in the valence band and vacancies in the conductance band.
Bottom: experimental PL spectra in $\sigma_{-}$ (blue) and $\sigma_{+}$ (red) polarizations measured in 35~nm QW at 4 T and 0.5 K.}
\end{figure}

At bath temperatures above 2 K, spin-triplet magnetoexcitons are ``ionized''~\cite{def:ionized}, and the recombination spectra exhibit two lines related to the transitions of photo-excited valence holes to the occupied states in the conductance band (Fig.~\ref{fig:fig1}). As the temperature drops, spin-triplet magnetoexcitons are accumulated, and two extra lines appear.
Let us consider feasible optical transitions from the zeroth Landau level of electrons to the zeroth Landau level of heavy holes.
The number of such transitions is not large since the spectrum of single-particle states is quantized, and the Fermi level lies in the middle of the cyclotron energy gap separating the zeroth and the first Landau levels.
Two obvious transitions are allowed in each of the two polarizations of the emitted photon (Fig.~\ref{fig:fig2}).
The first transfers the photoexited valence hole to the conductance band.
It is this transition that is observed at elevated temperatures.
The second occurs between inner states of a trion formed by a valence hole and a spin-triplet magnetoexciton.
This transition is dipole allowed if it does not change the internal quantum numbers of the trion~\cite{dzyubenko2000}.
The final state of the second transition in $\sigma_{-}$ polarization is a trion with vacancies in the zeroth Landau level having spins directed oppositely (Pln). In $\sigma_{+}$ polarization, it is a trion with vacancies having aligned spins (T).
The excited electron bound in the T-state is not able to jump to any of two available vacancies owing to spin conservation.
On the contrary, the excited electron bound in the Pln-state may jump to one of two available vacancies with emission of a cyclotron photon.
The latter may in turn be absorbed by the electrons in the zeroth Landau level to give birth to an electron-vacancy pair with the cyclotron energy, and so on.
If there were no extra vacancy bound into the spin-triplet magnetoexciton, the described quasi-particle would correspond to a magnetoplasmon with the cyclic quench and the reappearance of electron-hole pairs as plasma oscillations~\cite{kallin1984}. In the presence of the extra vacancy, the Pln-state is a \textit{bi-hole plasmaron}.

\begin{figure}
\includegraphics{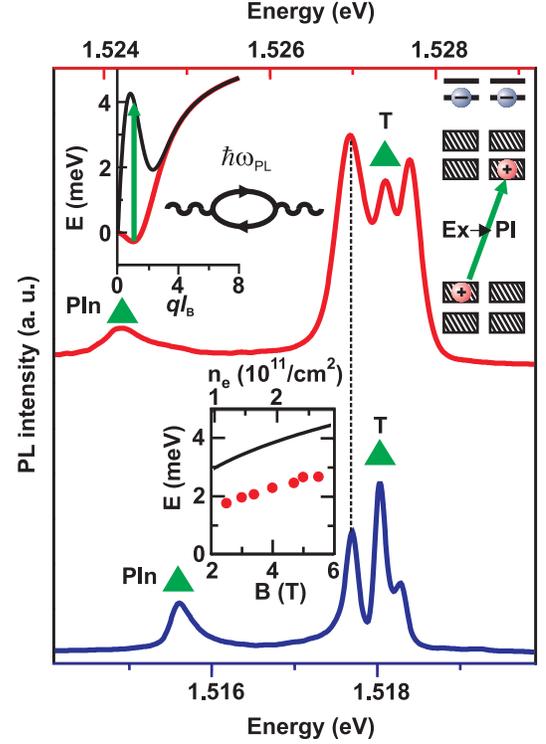}
\caption{\label{fig:fig3}
Top: PL spectra for narrow QW (17~nm, red curve) and broad QW (35~nm, blue curve). The equally scaled energy axes are shown in the corresponding colors.
The energies of single particle optical transitions in $\sigma_{-}$ polarization are aligned.
The top-right inset shows optical transition scheme used in the theoretical calculations.
The top-left inset shows the calculated exciton (Ex) dispersion curve (red) and magnetoplasmon (MP) dispersion curve (black).
The bottom inset shows the measured plasmaron energy shift (red dots) and the calculated magnetoplasmon energy shift (black line) vs. the electron density for 17~nm QW.}
\end{figure}

The expected optical transitions are observed in the recombination spectra (Fig.~\ref{fig:fig2}).
The energy of the inner transition for the T-trion almost coincides with the energy of single-particle optical transition at elevated temperatures.
The tiny energy shift of 0.2~meV is due to the difference in the Coulomb interaction among the particles comprising the T-trion as one of the holes passes from the valence to the conductance band.
The energy of the inner transition in the Pln-trion leading to the appearance of bi-hole plasmaron is significantly (2~meV) less than the energy of single-particle optical transition.
It reduces so dramatically owing to plasma oscillations.

An exact calculation of the plasmaron energy shift is an exceedingly complex problem~\cite{dzyubenko2000}.
To estimate the plasma oscilation energy, we consider a simplified model of optical transition from the initial two-particle state (exciton) to the final two-particle state (magnetoplasmon), excluding from consideration the vacancy bound in spin-triplet magnetoexciton (Fig.~\ref{fig:fig3}). The initial state is a $p$-type exciton (Fig.~3). The minimum of the exciton dispersion is near momentum $q_{m} \simeq 1 / l_B$ where $l_B = \sqrt{\hbar c / e H}$ is the magnetic length~\cite{kallin1984}.
The contribution to the final state energy made by the plasma oscillations is:
$$
E_{MP}(q_{m})=\frac{e^2}{\epsilon}q_{m}f(q_{m})e^{-\frac{(q_{m}l_B)^2}{2}},
$$
$f(q) = \int e^{-q |{z_1 - z_2}|} \Psi(z_1)^2 \Psi(z_2)^2 dz_1 dz_2 $ is the geometric form-factor arising as a result of the nonlocality of the electron in the growth direction of the quantum well ($z$) (in the strict 2D-limit $f(q)=1$). $\Psi(z)$ is the $z$-component of the electron wave function, and $\epsilon$ is the relative permittivity of GaAs (Fig.~3).
The calculated plasma oscillation energy $E_{MP}(q_{m})$ exceeds the measured plasmaron shift by about 30~\%.
This disagreement is not surprising considering the second-order corrections to the energies of the quasiparticles in question~\cite{kulik2005}, the complexity of the real valence hole wave function, and, finally, the influence of the extra vacancy.

\begin{figure}
\includegraphics{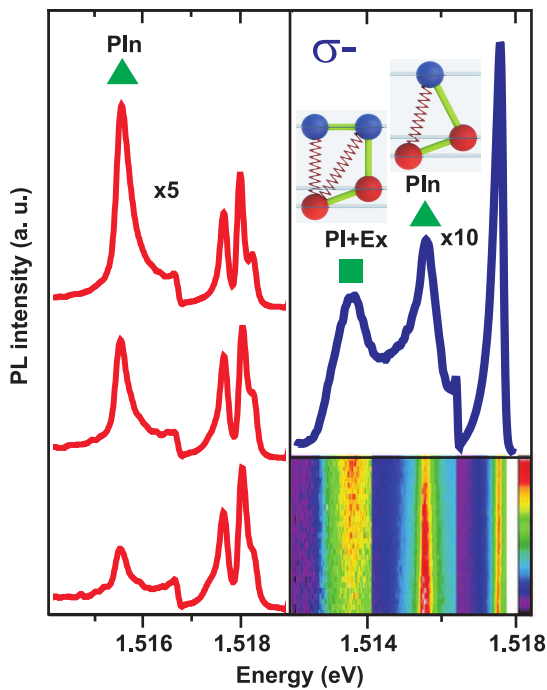}
\caption{\label{fig:fig4}
Left: PL spectra measured at different pumping powers of 1, 3, and 7.5~mW/cm$^2$, and the bath temperature of 0.5~K (from bottom to top).
Right: Bottom, PL spectra measured in pumping power range of 10-80~mW/cm$^2$ at temperatures of 0.5~K (the power increases in equal steps from bottom to top). Spectra are normalized to give equal intensities for three detected lines at 80~mW/cm$^2$.  Top: PL spectrum in $\sigma_{-}$ polarization measured at 80~mW/cm$^2$. The scheme demonstrates the corresponding final states.}
\end{figure}

The plasma origin for the Pln-energy shift is proved by measurements in two quantum wells of significantly different widths (a reduction in the well width enhances the plasma oscillation energy because of the geometrical form-factor $f(q)$)(Fig.~\ref{fig:fig3}).
In a more narrow QW, the plasmaron shift is larger by 1.4 times, which is exactly in accordance with the calculations. Further, the plasmaron shift measured in a wide range of electron densities agrees well with the calculated values, with the exception of the systematic shift to the lower energies (Fig.~\ref{fig:fig3}).
Thus, we establish the plasma nature of the energy shift for the Pln-line.
The cross-section for the plasmaron recombination increases when either the amount of spin-triplet magnetoexcitons or the extra electron-hole pairs increases (Fig.~4).
Yet, it saturates at the non resonant photoexcitation power density of 7.5 mW/cm$^2$. With a further increase in the photoexcitation power, the plasmaron recombination quenches and a new recombination channel with an energy shift twice that for plasmarons opens (Fig.~\ref{fig:fig4}).

The new recombination line Pl+Ex appears in $\sigma_{-}$ polarization of the emitting photon, which signals an optical transition of a hole from the valence band to the lowest spin sublevel of the conductance band (Fig~4). The only possible state one can construct of a spin-triplet exciton and an electron-hole pair apart from the plasmaron is the exotic four-particle state, {\it plasmon-exciton molecule}. The plasmon-exciton molecule forms at large photoexcitation densities when the probability to meet a photoexcited electron and a photoexcited hole simultaneously in the close proximity to a spin-triplet magnetoexciton  increases (Fig.~4). The formation of the plasmon-exciton molecule can be visualized as follows. In the initial state, a $p$-type exciton attaches to a spin-triplet exciton forming a bi-exciton molecule. When the hole bound in the bi-exciton transfers from the valence band to the conductance band conserving internal quantum numbers of the bi-exciton, a plasmon-exciton molecule forms. The corresponding recombination line is shifted owing to the plasma oscillations in the final state. In contrast to the plasmaron case, the plasmon-exciton molecule has two electrons both able to jump to the empty vacancy in the lowest spin sublevel (Fig.~4). This doubles the electron charge involved in the collective plasma oscillations. It is surprising that the plasma oscillations energy exactly doubles (within the experimental uncertainty) over plasmaron energy, notwithstanding the fact that two electrons and only a single vacancy in the electron Fermi sea participate in plasma oscillations (Fig.~4).

In conclusion, we created a super-long lifetime ensemble of non equilibrium excitations in 2D-metal under a quantising magnetic field from which possible three- and four-particle stable states involved in collective plasma oscillations are constructed. The stability of bi-hole plasmaron as well as an exotic four-particle state, \textit{i.e.}, a plasmon-exciton molecule, is demonstrated. The existence of a bound four-particle excited state in metals with collective plasma properties has not been conceived thus far, and hence, extensive theoretical efforts are needed to explain the experimental finding.

This research was partially supported by the Russian Foundation for Basic Research.

\end{document}